\documentstyle[twocolumn,epsf]{jpsj}
\def\runtitle{Numerical Study of the One-Dimensional Spin-Orbit Coupled System
with SU(2)$\otimes$SU(2) Symmetry}
\def\runauthor{Yasufumi {\sc Yamashita}, Naokazu {\sc Shibata} 
and Kazuo {\sc Ueda}}
\title{
Numerical Study of the One-Dimensional Spin-Orbit Coupled System\\
with SU(2)$\otimes$SU(2) Symmetry
}
\author{Yasufumi Yamashita, Naokazu Shibata$^1$,
and Kazuo Ueda}
\inst{
Institute for Solid State Physics, University of Tokyo,
Roppongi 7-22-1, Minatoku, Tokyo 106-8666, Japan\\
$^1$Institute of Material Physics, University of Tsukuba, 
Tsukuba 305-8573, Japan.}
\recdate{\today}

\abst{
We numerically study the SU(2)$\otimes$SU(2) symmetric spin-orbit coupled
model as a lower symmetric generalization of the SU(4) exchange model.
On the symmetric line with respect to the spin and orbit, 
our result shows the essentially singular gap formation 
in consistent with the analytic approach, 
which is different from the previous numerical calculation.
Furthermore, we find new critical phases around the SU(4) point,
surrounding the previously known gapless symmetric line.
In these novel phases either spin or orbital excitations around momentum $q=\pi$ 
form massless continua which split from 
the excitations belonging to the $[2^2]$ irreducible representation 
at the SU(4) point. On the critical symmetric line, 
the additional coupled spin-orbit excitations around $q=\pi/2$ 
originating from [$2^11^2$] become critical, too.
}
\kword{spin-orbit coupled system, orbital fluctuations, SU(4) symmetry, DMRG}
\begin{document}
\sloppy
\maketitle
\section{Introduction}
In connection with the study of the spin-orbit coupled systems with 
strong quantum fluctuations, the highest symmetric SU(4) exchange model has 
been a subject of the intensive study\cite{yamashita,FuChun,troyer}.
It describes the localized model of the orbitally twofold degenerate 
Hubbard model with the hoppings between the same type of the orbitals, 
neglecting Hund rule coupling.
The fundamental understandings of this most simplified model have been 
obtained beyond the scope of the mean-field analysis in one dimension.
In realistic materials, however, the highest SU(4) symmetry is almost 
always broken by an anisotropic hybridization or Jahn-Teller effect. 
Therefore, it is interesting as a next step to consider the physically 
relevant lower symmetric models around the SU(4) symmetric limit,
which may shed light on the physics of recent interest concerned with 
the spin and orbital fluctuations.
For example, we studied the SU(4) exchange model under magnetic fields 
with the resultant SU(2)$\otimes$U(1) symmetry and found the crossover 
phenomena between the SU(4) and the S(2) symmetric limit\cite{yamashita2}. 
As another example, a finite Hund coupling constant introduce the 
Ising-type anisotropy into the orbital part of the effective Hamiltonian, 
while the spin part keeps the SU(2) symmetry\cite{yamashita}.

As a different lower symmetric model, the following SU(2)$\otimes$SU(2) 
symmetric model has been studied by using the numerical and/or analytical 
method by several authors\cite{pati,azaria};
\begin{eqnarray}
H(a,b)=K\sum_{i}
\left(\vec{S}_{i} \vec{S}_{i+1}+\frac{a}{4}\right)
\left(\vec{T}_{i} \vec{T}_{i+1}+\frac{b}{4}\right),\label{model}
\end{eqnarray}
where $K$ is a positive constant and $\vec{S}$ and $\vec{T}$ are
the spin and orbital pseudo-spin 1/2 operators, respectively.
The study of this seemingly artificial model is stimulated by the recent 
experimental progresses in the quasi one-dimensional spin-gap 
materials such as ${\rm Na_2Ti_2Sb_2O}$ or ${\rm NaV_2O_5}$,
which are essentially described by the two-band Hubbard model at quarter 
filling\cite{pati}.
Kolezhuk $et\,al.$\cite{mikeska} studied $H(3,3)$ by using the matrix product 
ansatz and showed that this model has the doubly degenerate ground states 
with gapful excitations. 
Note that, in our definition, $H(1,1)$ is equal to the integrable SU(4) 
exchange model and gapless\cite{yamashita,FuChun}. 
Pati $et\,al.$\cite{pati} studied the model by using 
the density matrix renormalization group (DMRG) method, but their results 
are significantly different from ours in the most interesting region in the 
$a-b$ plane. On the symmetric $b=a$ line, 
our results are consistent with those of the recent 
analysis applying the bosonization method around the SU(4) 
symmetric point by Azaria $et\,al.$\cite{azaria}

The purpose of this paper is to present a renewed phase 
diagram over the entire $a-b$ plane, including newly found critical phases 
sandwiching the previously known gapless symmetric line. 
The four distinct phases centering the SU(4) symmetric point are
classified with the aid of the Lieb-Schultz-Mattis (LSM) theorem.
We are benefited from the Bethe-ansatz results of the three massless 
SU(4)-multiplet excitations and how they split away from the SU(4) point.
The concept of the simple renormalization group around the SU(4) fixed point
has also helps us to draw the gapless-gapful phase boundaries.

\section{Ground-state properties and Magnetic phase diagram}
\begin{figure}[hbt]\begin{center}
\hspace*{-1cm}
\leavevmode \epsfxsize=80mm 
\epsffile{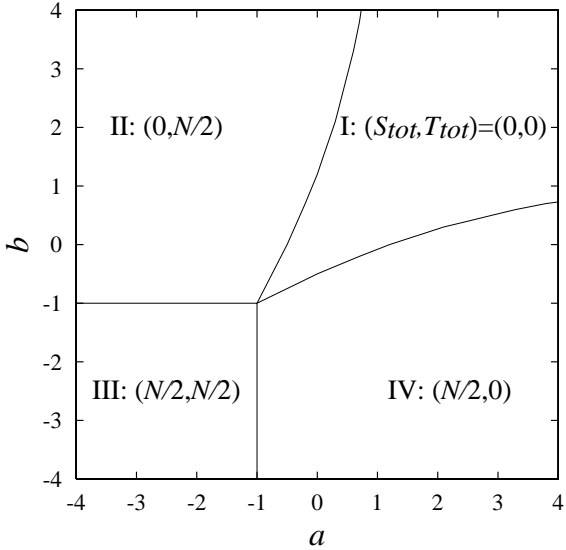}
\caption{Ground-state phase diagram of $H(a,b)$ with the PBC.
Each phase is classified according to the quantum number
$(S_{tot},T_{tot})$ as shown in the figure.}\label{abplane}
\end{center}\end{figure}

Let us begin our analysis by investigating the ground-state properties 
of $H(a,b)$ by using the Lanczos and DMRG method\cite{white}.
Though Pati $et\,al.$ have performed similar calculations,
their results are over-simplified. 
So we present more detailed ground-state phase diagram including
asymptotic behaviors and some exact proofs about these phases.
In what follows, we mainly use the periodic boundary 
conditions (PBC) for eliminating the boundary effects and consider the 
$N=4n$ number of sites to avoid the extrinsic effects due to the complex 
many-fold degeneracies\cite{yamashita}. 
The ground-states phase diagram in the two-dimensional $a-b$ plane is 
illustrated in Fig.\ref{abplane}. 
By the numerical calculations with changing the system 
sizes up to 20 sites, we have confirmed that the phase boundaries hardly move.

Besides the SU(4) symmetric point at $(a,b)=(1,1)$, where the model is 
Bethe-ansatz integrable, we can show that there are several regions in the 
$a-b$ plane where the ground states are determined exactly. In the region III 
($a\le -1$ and $b\le -1$), we can prove the following inequality; 
\begin{eqnarray}
&&H(a,b)=\sum_i 
\left(\frac{1}{4}-\vec{S}_i \vec{S}_{i+1}+|a_+|\right)
\left(\frac{1}{4}-\vec{T}_i \vec{T}_{i+1}+|b_+|\right)\nonumber\\
&&=\sum_i\left(P_{i,i+1}^{S=0}P_{i,i+1}^{T=0}+|b_+|P_{i,i+1}^{S=0}+
|a_+|P_{i,i+1}^{T=0}+a_+b_+\right)\nonumber\\
&&\ge \sum_{i}\left({a+1}\right) \left({b+1}\right)/16,
\end{eqnarray}
where $a_+=(a+1)/4$ and $b_+=(b+1)/4$, and $P_{i,i+1}^{S=0}$ 
($P_{i,i+1}^{T=0}$)
is the projection operator onto the $S$ ($T$) spin singlet for the neighboring
sites. The equality is satisfied when and only when both $S$ and $T$ spins
order ferromagnetically. Accordingly, the ground state is 
the complete ferromagnetic state with respect to both $S$ and $T$ spins.

In the part of phase IV ($a\ge-1$ and $b\le-1$), we can rewrite the 
$H(a,b)$ as follows;
\begin{eqnarray}
&&H(a,b)=\sum_i
\left(\frac{1}{4}-\vec{S}_i \vec{S}_{i+1}-a_+\right)
\left(\frac{1}{4}-\vec{T}_i \vec{T}_{i+1}+|b_+|\right)\nonumber \\
&&=\sum_{i}\left(
P_{i,i+1}^{S=0}P_{i,i+1}^{T=0}
+|b_+|P_{i,i+1}^{S=0}
-a_+P_{i,i+1}^{T=0}
-a_+|b_+| \right).
\end{eqnarray}
Completely ferromagnetic (FM) $S$ spins and disordered $T$ spins given by the 
$T$-spin 1/2 antiferromagnetic (AF) Bethe ansatz solution make the expectation
values of the second and third terms minimum at the same time. 
For these wave functions the first term is automatically minimized. Therefore
the ground states are constructed by such configurations as the $S$ and $T$ 
spin parts are given by FM-ordered state with $S_{tot}=N/2$ and the spin 
1/2 AF-disordered liquid state with $T_{tot}=0$. 
In fact, Fig.\ref{abplane} shows that these wave functions remain to be the 
ground states in a little wider region (still $a\ge -1$ but $b\ge -1$).

The asymptotic behavior of the I-IV phase boundary far away from origin is 
estimated through the consideration of the instability of the phase IV when
approaching this boundary from the exactly solvable region. 
To this end, we consider the infinitely large $a$ and finite
$b$ limits and apply the perturbative approach. The ground-state 
wave function for the non-perturbative term, 
including $a$, becomes the AF Bethe ansatz one about the $T$ spin. 
Then the perturbative term, not including $a$, have the form, 
$\sum_i(b/4-\epsilon_g)\vec{S}_i \vec{S}_{i+1}$, 
where $-\epsilon_g$ is the ground-state energy per site 
for AF Heisenberg model, defined by $\langle \vec{T}_i\vec{T}_{i+1} 
\rangle _{g.s.}$. Therefore, in the region with $b<4 \epsilon_g$ and 
large $a$, the ground-state wave function can be the same 
in the phase IV. 
We have checked that the I-IV phase boundary converges to
$b=4\varepsilon_g$ for the finite 8- and 12-site systems.
Since we know the exact value of $\epsilon_g=\log{2}-1/4$ 
in the bulk limit, we conclude that $b=4\varepsilon_g=1.7726$ is the 
asymptote of this phase boundary.
It is important to note that for the region $b>4\epsilon_g$, the initial 
wave functions with the decoupled $S$ and $T$ spin ones are no more valid
and such perturbative approach does not always work. According to the symmetry
of $H(a,b)$, in the phase II the model shows the same 
properties as in the phase IV with changing the $S$ and $T$ spins. 

At $(a,b)=(3,3)$, we can rewrite $H(a,b)$ as
\begin{eqnarray}H(a,b)=\sum_{i}P_{i,i+1}^{S=1}P_{i,i+1}^{T=1},\end{eqnarray}
where $P_{i,i+1}^{S=1}$ ($P_{i,i+1}^{T=1}$) represents the projection operator
onto the $S$ ($T$) spin triplet subspace for the neighboring sites. 
The configuration where either $S$ or $T$ spins are singlet at every bond
makes the energy minimum to be zero. 
Thus we find that the nested dimer singlet states, 
where $S$ and $T$ spin parts are dimerized so as not 
to overlap their singlet bonds, are the exact ground states\cite{mikeska}. 
These states are obviously twofold degenerate and the relative difference of 
the momentum equals to $\pi$. 
Since the LSM theorem\cite{LSM} is applicable to 
the present model for arbitrary $a$ and $b$, there exists a 
finite gap above these degenerate ground states.
Lastly, in the infinitely large $a$ and $b$ limit, 
Eq.(\ref{model}) was 
studied as the two-leg spin ladder model with 4-body interaction and 
found to have a finite gap\cite{tsvelik}.

\section{Excitations and new gapless phases}
\subsection{Excitations in the phase II, III, and IV}
\begin{figure}[tb]\begin{center}
\hspace*{-1.8cm}
\leavevmode \epsfxsize=80mm \epsffile{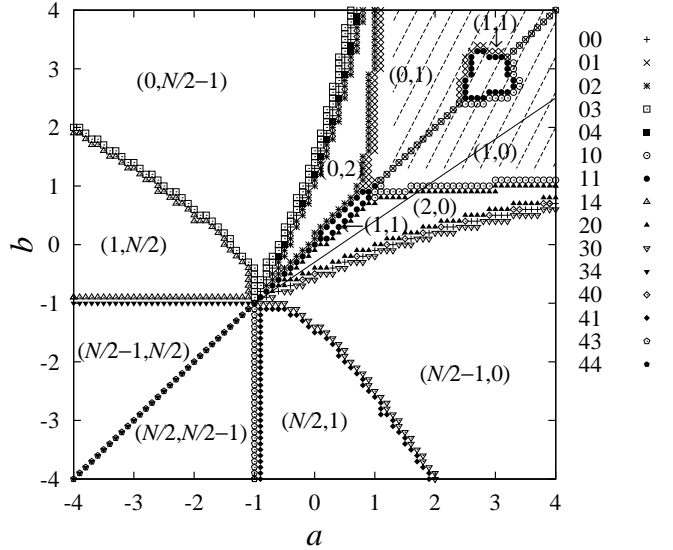}
\caption{The excited-state phase diagram for the system size $N=8$ 
with the PBC. The right-hand side symbols stand for the quantum number 
$(S_{tot},T_{tot})$. 
The region (2,0) continues to about $(a,b)=(7.2,1.2)$ with decreasing its 
width. In the shaded region, the ground states are nearly twofold degenerate.
}\label{abplaneex}\end{center}\end{figure}

Next we investigate properties of the excitations in the $a-b$ plane. 
For this purpose, we classify the excitations according to their quantum 
numbers $(S_{tot},T_{tot})$. 
The several lowest-lying energies of $H(a,b)$ are calculated by using the 
Lanczos method in all the subspaces with changing $(S_{tot}^z,T_{tot}^z)$. 
Note that there are $N/2 \times N/2$ number of 
subspaces for the $N$-site system.
Through these calculations we can specify the quantum number 
$(S_{tot},T_{tot})$ of the excited state at a point $(a,b)$. We have 
performed above calculations in the square region, $-4 \le a,b \le 4$, with 
varying $a$ and $b$ by the mesh of 0.1. As a result, we obtain the 
$80\times 80$ meshes, on each mesh the quantum number $(S_{tot},T_{tot})$ 
of the lowest-lying excited state is specified. The obtained boundaries of 
the excitations for the 8-site system are shown in Fig.\ref{abplaneex}.
In the figure, we label the states so as to be invariant for
the general system size $N$. The spin or orbital
excitations with seemingly strange quantum numbers (2,0)
or (0,2), respectively, are intrinsic independently of $N$.
 
In the phase III, the usual FM-magnon excitations of either $S$ or $T$ spin 
degrees of freedom are observed depending on $a<b$ or $a>b$, respectively.
These excitations are exactly consistent with those expected from the 
completely FM ground state. In the phase IV, both FM $S$-spin magnon 
excitations and AF $T$-pseudo-spin magnon excitations appear next to 
each other. This boundary can be understood in the following manner.
First of all, we note that the AF magnons have a 
momentum($q$)-linear dispersion while the FM magnons have a $q$-square 
dispersion. Therefore the FM-magnon excitation always gives lower energy 
than the AF magnon for a sufficiently small momentum. 
Unless the system size is large enough and the minimum discrete momentum is 
not smaller than the crossing point of the FM- and the AF-magnon dispersion, 
the first 
excited state is the AF magnon. In the bulk limit, however, the FM magnon 
necessarily gives the minimum excitation. In fact, the numerically calculated 
region of the AF $T$-pseudo-spin magnon excitations are suppressed 
with increasing the system size. 
Therefore we conclude that the lowest excitation 
in the phase IV is the FM $S$-spin magnon in the bulk limit. 
By exchanging $S$ and $T$ spins in the region IV, 
we know the excitation properties in the phase II. 

The boundary between (2,0) and ($N/2-1$,0) in Fig.\ref{abplaneex} is 
compatible with the phase boundary between I and IV of the ground-state 
phase diagram (Fig.\ref{abplane}). On this boundary, 
there is the narrow region where the lowest excitation has either
$(S_{tot},T_{tot})=$ (0,0) or ($N/2$,0), which are the quantum numbers 
of the neighboring ground states. This is because magnetic and
non-magnetic ground state cross on it and the energies of the ground states on
the other side phase become lower than those of the magnon excitations owing 
to the finite size effects. From the arguments until now, the ground state 
and excitations in the phases II, III, and IV, are clearly understood. 
It should be noted that above simple pictures come from the decoupled
wave functions about spin and orbital degrees of freedom.
The remaining problems are the properties in the phase I, 
especially nearly around the SU(4) symmetric point 
where the decoupled spin-orbit picture is obviously wrong,
which are the subjects in the rest of this paper.

\subsection{Symmetric line in the phase I}
\begin{figure}\begin{center}
\leavevmode
\epsfxsize=70mm
\epsffile{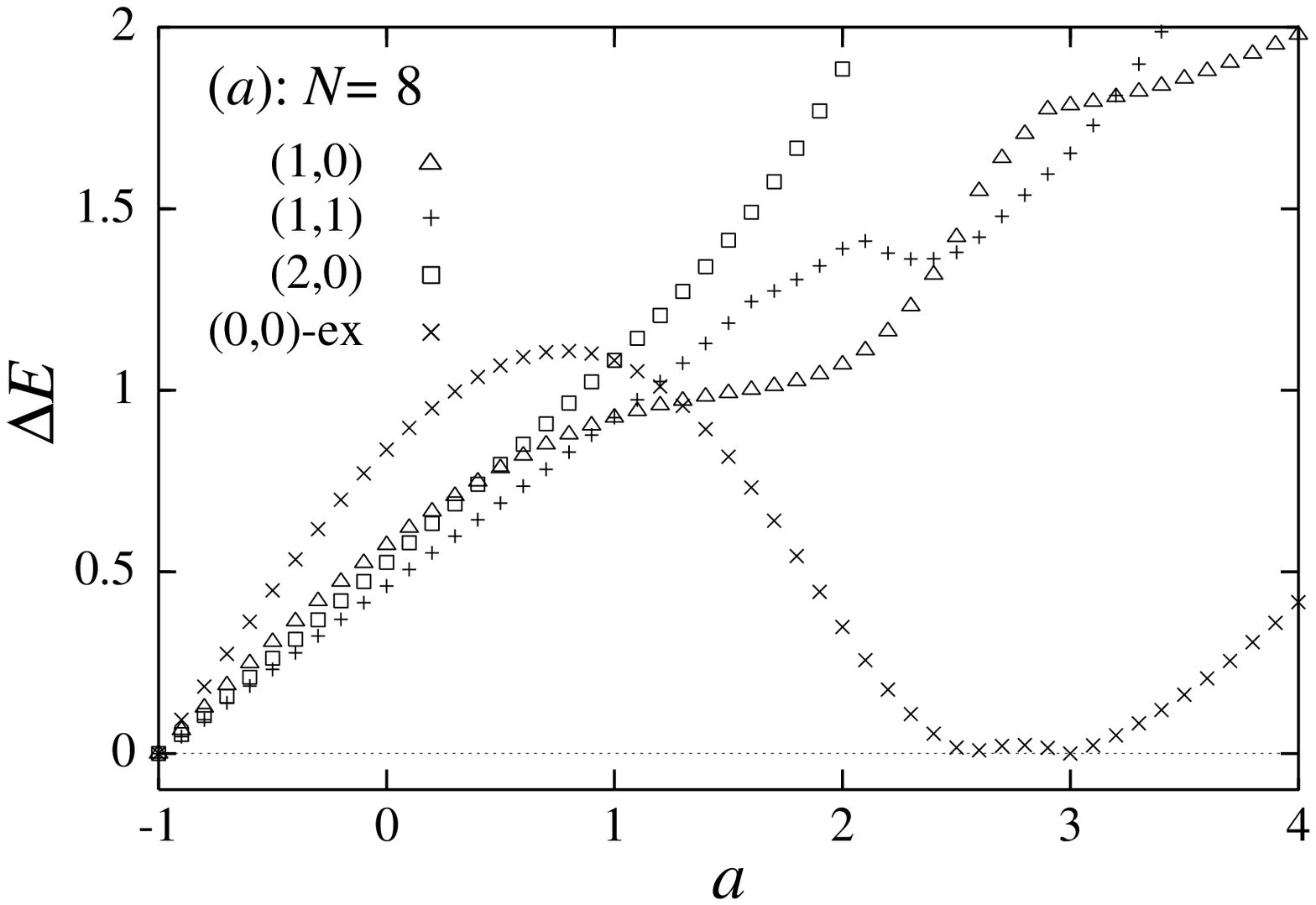}\\ \noindent
\epsfxsize=70mm
\epsffile{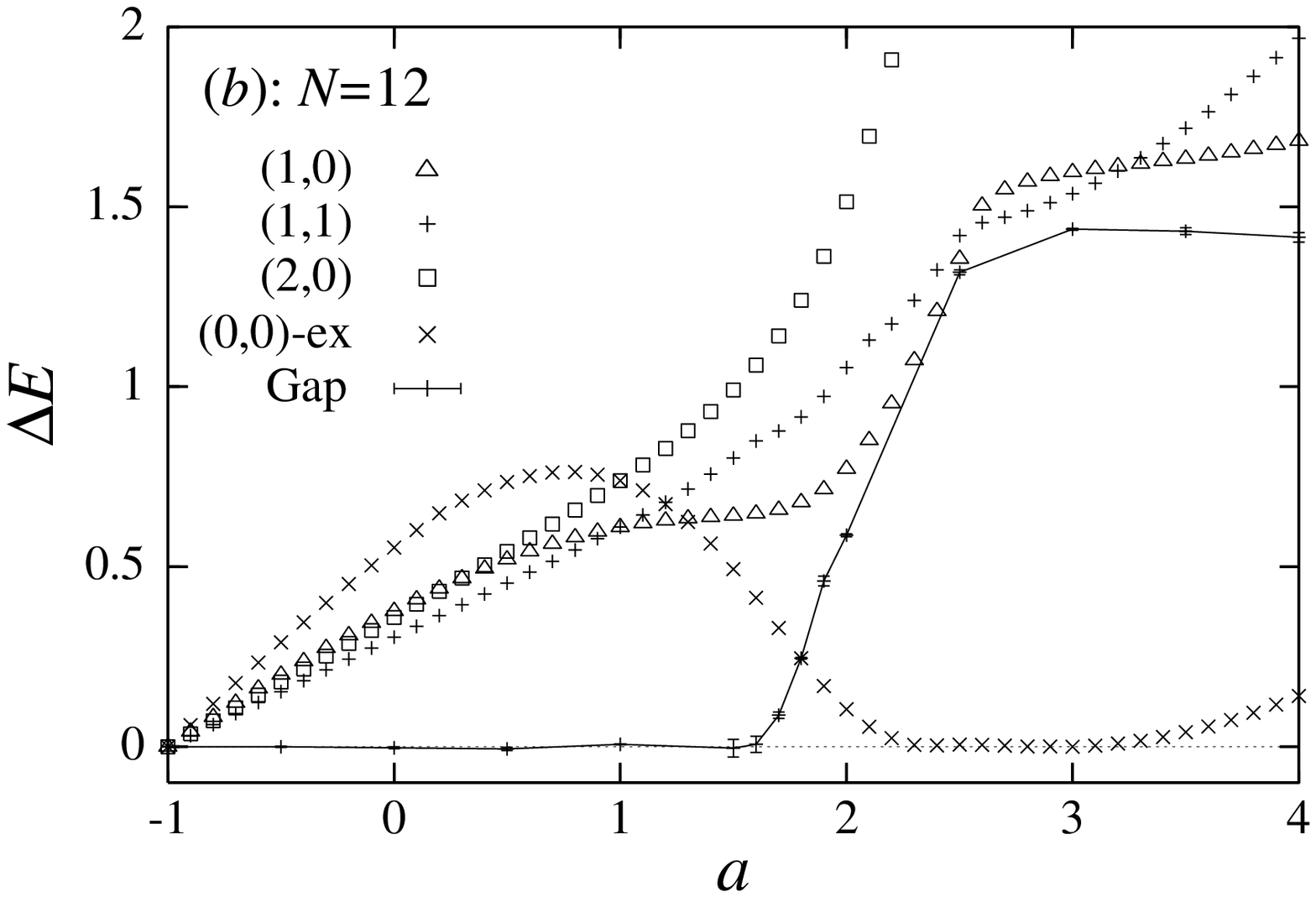}
\caption{ 
The excitation-energy diagram 
with the specified quantum numbers $(S_{tot},T_{tot})$
on the symmetric $b=a$ line.
Figure (a) and (b) correspond to 8- and 12-site system, respectively.
(0,0)-ex represents the excited state with $(S_{tot},T_{tot})=(0,0)$.
Figure (b) also includes the excitation gap
after the system-size extrapolation.
}\label{a=b8_12}
\end{center}\end{figure}

It is naturally expected that the SU(4) symmetric point becomes the 
multi-critical fixed point and 
how different phases emerge from it determine the character of each phase.
We would like to emphasize that one of the main 
concerns hereafter is the fact whether a finite gap exists or not above the 
ground state in the thermodynamic limit. First, in order to consider the 
properties on $b=a$ line (symmetric line), which has relatively higher 
symmetry in the phase I.
We keep track of the several lowest eigenstates with the specified quantum 
numbers $(S_{tot},T_{tot})$ by using the exact diagonalization. The obtained 
excitation-energy diagrams for the 8- and 12-site system with the PBC are 
shown in Fig.\ref{a=b8_12}. 
At $a=1$, we know that $H(a,a)$ has gapless excitations with 
$(S_{tot},T_{tot})=(1,1)$, (1,0), (2,0), and others\cite{yamashita}.
Similarly the states with these quantum numbers are softened at $a=-1$.
From Fig.\ref{a=b8_12} the excitation energies ($\Delta E$) 
seem to be uniformly 
suppressed in the interval of $-1<a<1$ as the system size is increased.
Thus it is natural to consider that the symmetric line is critical 
in the range of $-1\le a\le 1$. We have confirmed 
that the excitation energies corresponding to $(S_{tot},T_{tot})=(1,1)$ and 
(2,0), (=(0,2)), all goes to zero linearly by the DMRG calculations 
for the system size up to 20 sites at $a=-0.5$, 0, 0.5, and 1.
The obtained excitation gap after the system size extrapolation 
are presented in Fig.\ref{a=b8_12}-(b), in which
we fit the finite-size data with the polynomial function of $1/N$ 
by the least mean 
square method. In the DMRG calculations, we keep at least 400 number of states
in each renormalization. 

This gapless behavior is consistent with the numerical calculation by 
Pati $et\,al.$ and the analytic approach by Azaria $et\,al.$
On this line, we perform the momentum-specified Lanczos method using
translational symmetry and specify the (1,1)- and (2,0)- excitation
to have the momentum $\pi/2$ and $\pi$, respectively. From the LSM theorem, 
these excitations must not be the isolated states but 
form continua. 
As a matter of fact, there are three independent massless continua
with the same velocity of sound at the SU(4) point\cite{sutherland}.
Both the lowest and highest of them, softening at $q=\pi/2$, belong to
$[2^11^2]$ irreducible representation, consisting of 
$(S_{tot},T_{tot})=(0,1)\oplus(1,0)\oplus(1,1)$. 
On the other hand, the middle one softening at $\pi$ belongs to 
$[2^2]=(0,0)\oplus(1,1)\oplus(2,0)\oplus(0,2)$\cite{yamashita}.
Therefore, we conclude that, along the symmetric line, (1,1)- and (2,0)- 
(same for (0,2)-) excitations are split from the [$2^11^2$] and [$2^2$] 
irreducible representations at the SU(4) point, respectively.

For $a>1$, the (0,0)-excitation with momentum $\pi$ falls down just
after the level crossing around $a=1$. At the SU(4) point we know that this 
state belongs to [$2^2$] irreducible representation
and its excitation energy is of the order of the inverse of the system size.
Therefore the crossing points of the lowest branches are expected to
converge to the SU(4)
point in the bulk limit. This is perhaps because that this point 
becomes the fixed point in the sense of the renormalization group.
For the region $a>1$, two lowest states with the same quantum number (0,0)
become nearly degenerate as the increase of $N$ shown in Fig.\ref{a=b8_12}.
In particular, at $a=3$ they completely degenerate even in the 
finite system sizes. Around there (1,1)-excitation becomes the lowest 
(see Fig.\ref{abplaneex}), which may be the spin-orbit bound states.

From these evidences,
it is reasonable to conclude that in the range of $a>1$ on the symmetric line,
the ground states are doubly degenerate breaking the translational symmetry. 
One of them corresponds to the ground state with $[1^4]$ irreducible 
representation and the other is the isolated state originating from $[2^2]$
at the SU(4) point.
Since the LSM theorem is applicable all over the $a-b$ plane as mentioned 
earlier, the doubly degenerate ground states mean the existence of the finite
excitation gap above them in the bulk limit,
while the other case of the LSM theorem, the singlet ground state with
continuous excitation spectrum, are realized for $a\le1$.

In order to study how the excitation gap opens along the symmetric line, 
the DMRG calculation for larger systems under the PBC is performed.
We estimate the first excitation energy by the difference between the two
lowest states in the subspace of $(S_{tot}^z,T_{tot}^z)=(0,0)$ and (1,0),
respectively.
The excitation gaps for $a>1$ calculated in the same way for $a\le1$ are shown 
in Fig.\ref{a=b8_12}-(b).
In the region of $a\ge 1.7$ there seems to exist a finite gap, but it is hard 
to conclude whether a finite gap exists or not in the region $1\le a \le 1.6$ 
even by the DMRG calculations up to 20-site system. From these results we 
conclude that for $a>1$ the excitation gap open exponentially slowly, while 
for $a\le1$ no finite gap exists. 

Let us consider a possible scenario by using the simple renormalization 
group concepts, in which we can explain the numerically obtained
excitation properties along the symmetric line. Suppose that we 
introduce the perturbation along this line into the SU(4) symmetric 
point, $a=1$. Then consider what kinds of the properties, such as 
relevant, irrelevant or marginal, the introduced perturbation should have 
so as to reproduce the numerical results.
It is likely that the dominant perturbation operator becomes a marginally 
irrelevant in the (-1,-1) direction, while it becomes a marginally relevant 
in the opposite direction. This picture is consistent 
with the numerical results, that is, for $a\le 1$ there does not exist a 
finite gap and for $a\ge 1$ a gap opens exponentially slowly. This scenario 
also agrees with the bosonization result by Azaria $et\,al.$ 

On the contrary, Pati $et\,al.$ claimed that the excitation is gapless 
for $a\le1$, though for $a>1$ the gap ($\Delta E$) 
opens by a power-law function, $\Delta E \propto (a-1)^{1.5}$.
They assumed that the first excited state was in the subspace of
$(S_{tot}^z,T_{tot}^z)=(1,1)$. 
However, such assumption does not give correct lowest-excitation energy
in almost all the region I except for the $b=a$ line $(-1\le a \le 1)$ and 
the neighborhood of $(a,b)=(3,3)$ (see Fig.\ref{abplaneex}). 
Their conclusion is unlikely from the viewpoint of the simple 
perturbative discussion.
The gapless behavior for $a<1$ requires irrelevant or marginally
irrelevant perturbative operators. Since the 
additional opposite-signed perturbation has the same scaling dimension,
the excitations in the opposite (1,1)-direction can not be gapful
with a power-low dependence.

Along the symmetric line, we have also calculated the correlation functions
by the DMRG method under open boundary conditions up to 36 sites.  
For eliminating the boundary effects, we adopt the averaging procedure and 
obtain their fourier transformations by arranging them in a ring. 
The peak structures exist at $q=\pi/2$ for $1<a\le2.0$, though
the peaks lose their sharpness and become broadened over $a>1.6$.
Note that, a large finite-size effect with 
the slow growth of the excitation gap make it difficult to determine the 
transition point ($a=1$) by this approach, too. For the region $a>2.1$, 
the peaks locate at $q=\pi$ and no other structures are found elsewhere.
The incommensurate line from $a=1$ to 2 discussed by Pati $et\,al.$ is not 
found. Our numerical result again supports the analysis by Azaria $et\,al.$

\subsection{Novel gapless phases in the asymmetric regions of the phase I}
\begin{figure}[htb]\begin{center}
\leavevmode
\epsfxsize=70mm
\epsffile{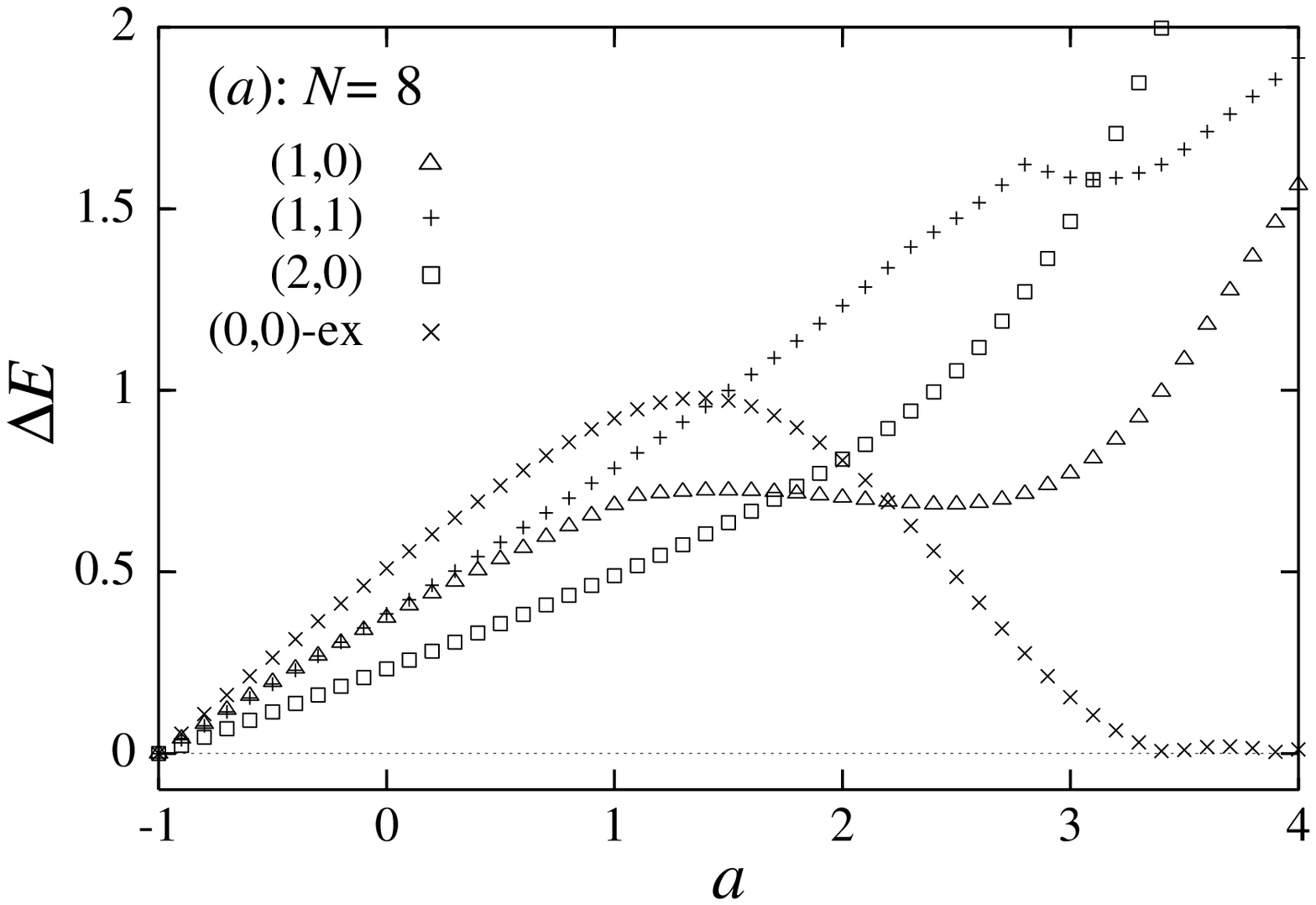}\\ \noindent
\epsfxsize=70mm 
\epsffile{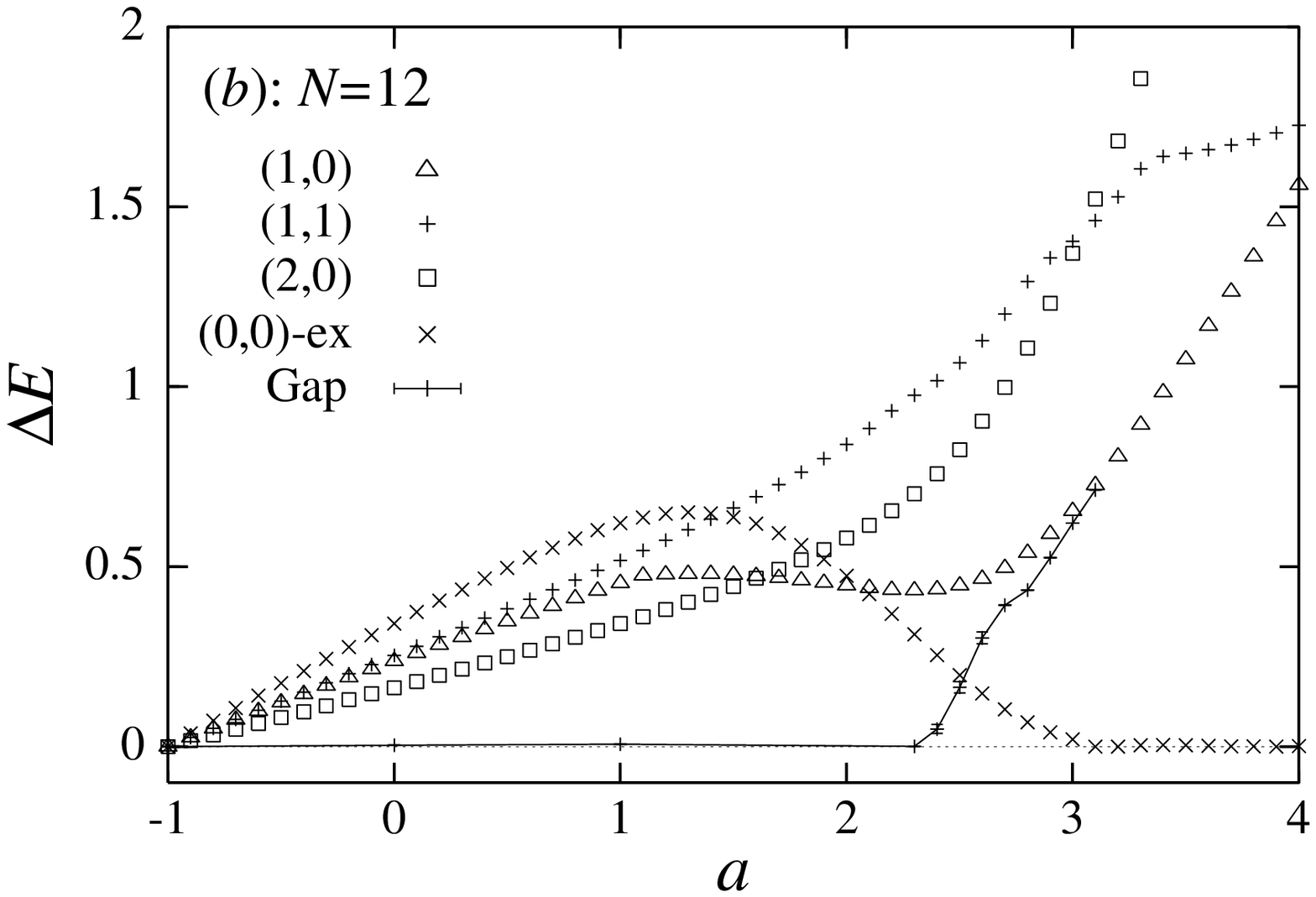}
\caption{The excitation-energy diagram 
on the asymmetric $b=0.7a-0.3$ line. The notations are the same as
Fig.3.}
\label{a=b8_12_asy}
\end{center}\end{figure}

Beside the symmetric critical line, we discover the novel distinct phases 
characterized by the strange excitation-quantum numbers, 
$(S_{tot},T_{tot})=(2,0)$ or (0,2), as shown in Fig.\ref{abplaneex}.
In order to investigate these phases, the excitation-energy diagrams 
are examined for 8- and 12-site system with PBC along the asymmetric 
$b=0.7a-0.3$ line, which is shown in Fig.\ref{abplaneex} by the solid line. 
The obtained diagrams, Fig.\ref{a=b8_12_asy}-(a) and (b), 
are different from those
on the symmetric line in that the level-crossing point is at about $a=1.9$
and below it the (2,0)-excitation becomes the lowest.
The lowest (2,0)-excitations with momentum $\pi$ are continuously 
connected to those on the symmetric line and therefore 
form the continuous excitation spectra softening at $q=\pi$. 
The lowest excitation gap goes to zero 
linearly at $a=0$ and 1 on $b=0.7a-0.3$ line by the DMRG calculation up to 
20-site system. 
The other gapless excitation with ($S_{tot},T_{tot})=(1,1)$ on the symmetric 
line also seems to have zero gap from the DMRG calculation up to system size 
with $N=20$ at least $a=1$.
The finite excitation gaps are obviously established over $a>3$, 
since even For $N=8$ and 12 sites the excitation energies converge as the 
function of $1/N$ as shown in Fig.\ref{a=b8_12_asy}-(a) and (b).

By using the DMRG method under the PBC, we research how the excitation 
gap opens along the asymmetric line from $a=-1$ to 4. The calculated 
conditions are the same for the symmetric line and the obtained results 
are shown in Fig.\ref{a=b8_12_asy}-(b). We find that the excitation gap opens 
exponentially slowly just after the sudden 
decrease of the (0,0)-excitations to be one of the degenerate ground states
in the bulk limit.
We can easily understand the behaviors of the excitation gap by looking at
those of the (1,0)-excitations in Fig.\ref{a=b8_12} and Fig.\ref{a=b8_12_asy}.
In short, all these gap-formative excitations show the exponentially slow 
growths after the level-crossing around $a=1$ or $1.9$.
Accordingly, the newly found (2,0)- and (0,2)-phases are gapless and 
in crossing the gapless-gapful phase boundary in the phase I, 
the gap opens exponentially slowly as well as on the symmetric line.

We have confirmed by the Lanczos method for 12-site system
that the more downward the boundary between (2,0)- and (1,0)-phases moves, 
the farther we are away from the SU(4) point.
By the DMRG calculation up to the system size $N=20$,
the lowest excitation energy at (3,0.8) in the subspace of 
$(S_{tot}^z,T_{tot}^z)=(1,0)$ is lower than that in (2,0).
On the other hand, those at (1,0.4) and (2,0.6) are involved in (2,0)-phase
and go to zero linearly as the function of $1/N$.
Our numerical results suggest that the gapless-gapful phase boundary 
in the phase I starts from the SU(4) point with some finite angle less than
right angle against the symmetric line, having a positive curvature.
It nearly converges to the ground-state phase boundary over $a\approx 3$.
Although we cannot numerically determine its exact angle,
the previously used simple picture based on the 
renormalization group tells us a reasonable scenario as follows. 
The key point is that
it is presumably expected that
when the perturbative operator in one direction are marginally irrelevant, 
that in the opposite direction will become marginally relevant.
Since no other phases are found in both gapless and gapful phases numerically, 
it is natural to suppose that the boundary starts
from the SU(4) point perpendicularly to the symmetric line,
so as not to contradict the symmetry about the exchange of the $S$
and $T$ spins
We think that our numerical results are naturally explained in accordance 
with this scenario. 

It should also be mentioned here that other possibilities still remain 
at this stage, since it is too difficult to distinguish a true zero gap
from a finite but exponentially small gap.
The analytic approach around the fixed point including the asymmetric
regions just meets this demand. Such analytic approach by using the 
bosonization method is now being developed by Y. Tsukamoto and 
N. Kawakami\cite{tsukamoto}. 
Their result seems to consistent with ours about the properties of 
each phase. It also consistently suggests the gapless-gapful boundaries 
begin exactly in the vertical direction to the 
symmetric line just around the SU(4) fixed point.

\begin{figure}[htb]\begin{center}
\hspace*{-1.6cm}
\leavevmode
\epsfxsize=75mm
\epsffile{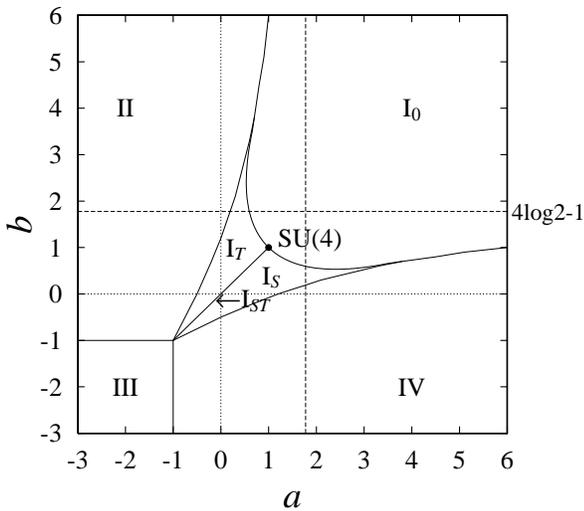}
\caption{
The phase diagram of SU(2)$\otimes$SU(2) coupled spin-orbit model given by 
Eq.(1) in the bulk limit. The region I in Fig.1 is decomposed into 
four phases, I$_0$, I$_S$, I$_T$, and I$_{ST}$.
The broken lines represent the asymptotes of the I-IV and I-II boundary
}\label{lastphase}
\end{center}\end{figure}

From the discussions above, we conclude that the intrinsic phase diagram
of the SU(2)$\otimes$SU(2) model is given by Fig.\ref{lastphase}.
Although there still remain some ambiguities about the I$_0$-I$_S$ and 
I$_0$-I$_T$ phase boundaries. 

\section{Summary}
In summary, we have studied the one-dimensional SU(2)$\otimes$SU(2) 
symmetric spin-orbit coupled model by means of the numerical calculations 
using the Lanczos and the DMRG methods. 
In the phase II, III, and IV, we have made clear the ground-state and 
excitation properties by noticing the fact that the ground-state wave 
functions are depicted by the decomposed $S$ and $T$ spins.
In addition we also determine the asymptote of phase I-IV boundary through the 
discussion about the instability of the phase IV.
On the symmetric line in the phase I, 
there exists a finite excitation gap for $a>1$, which grows exponentially 
slowly above the doubly degenerate ground states in consistent with the
LSM theorem.
On the other hand for $a<1$, the other case of the LSM theorem are realized
with the continuous spin, orbit, and coupled spin-orbit excitations 
corresponding to the excitation-quantum numbers,
$(S_{tot},T_{tot})=(2,0),(0,2),$ and (1,1), respectively.
We find that the estimated gap growth disagrees with that by Pati $et\,al.$
and reproduce the result by Azaria $et\,al.$

We extend our analysis in the asymmetric regions and find that the phase I is 
composed of the four distinct phases centering the SU(4) symmetric point
(see Fig.\ref{lastphase}). 
Three massless phases, I$_S$, I$_T$, and I$_{ST}$, are characterized 
by the gapless continua, above a singlet ground state, with the quantum 
number, $(S_{tot},T_{tot})=(2,0)$, (0,2), and (1,1), respectively.
Considering the continuous branches with the highly degenerate irreducible 
representations at the SU(4) point, we conclude that 
the pure spin or orbital excitations originate from the same branch softening 
at $q=\pi$ with the [$2^2$] irreducible representation,
while the coupled spin-orbit excitation from the other one 
softening at $q=\pi/2$ with [$2^11^2$] .
In the rest part of I, labelled as I$_0$,
the finite excitation gap opens above the doubly
degenerate ground states by breaking translational symmetry.
These gapless and gapful situations exactly correspond to the alternative
of the LSM theorem.
We also estimate 
how the excitation gap opens in crossing the gapless-gapful boundary 
in the phase I 
to find 
that the gap opens exponentially slowly in the same way on the symmetric line.

\section*{Acknowledgements}
The authors wish to acknowledge Norio Kawakami and Yasumasa Tsukamoto 
for many valuable discussions.
This work is financially supported by a Grant-in-Aid for Scientific 
Research from the Ministry of Education, Science,
Sports and Culture in Japan.

\end{document}